\documentclass[
reprint,
reprint,
nofootinbib,
 amsmath,amssymb,
 aps,
prr,
]{revtex4-2}
\usepackage[utf8]{inputenc}
\usepackage{listings}

\usepackage{graphicx} 
\graphicspath{{figures}}

\usepackage{tikz}
\usetikzlibrary{quantikz2}

\usepackage{xcolor}
\usepackage{graphicx}
\usepackage{dcolumn}
\usepackage{bm}
\usepackage{bbm}
\usepackage{amsmath}

\usepackage{orcidlink}
\usepackage{physics}
\usepackage{natbib}
\usepackage[version=4]{mhchem}
\usepackage{gensymb}
\makeatletter
\let\old@makecaption=\@makecaption
\usepackage{subcaption}
\let\@makecaption=\old@makecaption
\makeatother
\usepackage{siunitx}
\usepackage{graphicx,multirow}
\usepackage{array}
\usepackage{hyperref}
\hypersetup{
    colorlinks,
    linkcolor={red!80!black},
    citecolor={blue!80!black},
    urlcolor={blue!80!black}
}
\usepackage[capitalize]{cleveref}
\crefname{section}{Sec.}{Secs.}
\usepackage[super]{nth}
\usepackage{lipsum}

\usepackage{dsfont}

\begin{document}
\title{Classical optimization of reduced density matrix estimations with classical shadows using N-representability conditions under shot noise considerations}
\date{\today}

\author{Gian-Luca R. Anselmetti\orcidlink{0000-0002-8073-3567}}
\email{gian-luca.anselmetti@boehringer-ingelheim.com}
\affiliation{Quantum Lab, Boehringer Ingelheim, 55218 Ingelheim am Rhein, Germany}

\author{Matthias Degroote\orcidlink{0000-0002-8850-7708}}

\affiliation{Quantum Lab, Boehringer Ingelheim, 55218 Ingelheim am Rhein, Germany}

\author{Nikolaj Moll\orcidlink{0000-0001-5645-4667}}
\affiliation{Quantum Lab, Boehringer Ingelheim, 55218 Ingelheim am Rhein, Germany}

\author{Raffaele Santagati\orcidlink{0000-0001-9645-0580}}
\affiliation{Quantum Lab, Boehringer Ingelheim, 55218 Ingelheim am Rhein, Germany}

\author{Michael Streif\orcidlink{0000-0002-7509-4748}}
\affiliation{Quantum Lab, Boehringer Ingelheim, 55218 Ingelheim am Rhein, Germany}

\begin{abstract}
Classical shadow tomography has become a powerful tool in learning about quantum states prepared on a quantum computer. Recent works have used classical shadows to variationally enforce N-representability conditions on the 2-particle reduced density matrix. In this paper, we build upon previous research by choosing an improved estimator within classical shadow tomography and rephrasing the optimization constraints, resulting in an overall enhancement in performance under comparable measurement shot budgets. We further explore the specific regimes where these methods outperform the unbiased estimator of the stand-alone classical shadow protocol and quantify the potential savings in numerical studies.
\end{abstract}
\maketitle

The simulation of quantum chemistry presents a hard challenge for the field of computation. In recent years, quantum computation has been proposed to aid quantum chemistry, offering to simulate complex systems that previously were out of reach for conventional methods~\cite{Abrams_1997, Aspuru_Guzik_2005, Cao_2019, McArdle_2020}. However, current quantum computers are still limited in size and the number of operations due to noise and the absence of error correction~\cite{Preskill_2018}. As such, stand-alone quantum algorithms competing with the highly optimised methods of traditional quantum chemistry are still out of reach for the foreseeable future. This has opened up an avenue of research at the intersection of purely classical methods with the support of newly proposed simulation techniques on a quantum computer~\cite{Huggins_2022, scheurer23:_tailor_exter_correc_coupl_clust_quant_input}.

Early efforts have sought to combine the strengths of classical and quantum algorithms in an approach labelled as hybrid quantum computation. In this spirit, recent work has tried to enhance some established classical methods with data gathered from shadow tomography, a powerful tool preparing efficient classical representations of quantum states prepared on a quantum computer for certain observables~\cite{Huang_2020,zhao20_fermion_partial_tomog_via_class_shadow,Wan_2022,low22_class_shadow_fermion_with_partic_number_symmet}. An example is the combination of this tool with traditional methods of quantum chemistry like Quantum Monte Carlo (QMC)~\cite{Huggins_2022} and Coupled Cluster (CC)~\cite{scheurer23:_tailor_exter_correc_coupl_clust_quant_input}, achieving improvement in accuracy in certain situations. These techniques are motivated by the fact that learning on quantum data in the form of classical shadows is known to provide quantum advantage~\cite{Huang_2022} under certain conditions, identifying sampling on a quantum computer as a minimal-cost useful task.

While the aforementioned methods have focused on enhancing classical methods with quantum data, other work has sought to apply techniques developed in classical quantum chemistry to quantum computation, e.g., constraints stemming from the N-representability conditions (limiting the space of valid reduced density matrices (RDM) that can stem from a physical wave function) to enhance quantum readout~\cite{Rubin_2018,Hartree2020,avdic23:_fewer_n}. The goal of these works is to optimise measurement methods in sampling from an unknown quantum state, mostly trying to decrease sample variance and reduce sensitivity to device error~\cite{Rubin_2018,Hartree2020} and to decrease the shot budget of shadow tomography~\cite{avdic23:_fewer_n} with the use of v2rdm~\cite{iii23:_variat_deter_two_elect_reduc_densit_matrix,VERSTICHEL20112025}, a method casting the N-representability conditions into a semidefinite program. To achieve industrial relevance, extracting ground-state energies from quantum calculation does not lead to an application per se. The promise lies more in the efficient extraction of further chemically relevant observables, like e.g., molecular forces for geometry optimizations or molecular dynamics~\cite{Simon2024,Simon2024Amplified}, as this is paramount to compete with established classical methods~\cite{Santagati2024perspective}. As fault-tolerant algorithms require a large overhead to extract expectation values of observables on ground states~\cite{O_Brien_2022,Steudtner_2023}, sampling shadows of ground states from a quantum computer could help extract reduced density matrices more efficiently from a variational 2-RDM (v2rdm) procedure.

In this Letter, we introduce a new way of constraining the semidefinite program by using an improved classical shadow estimator of the entire 2-RDM estimate. We also further explore the role of shot noise in estimation problems with and without the use of v2rdm in the classical shadow tomography. We establish regimes where the method can lead to savings in shot budgets of up to a factor of 15 over the non-optimised estimate. We do not find that previous approaches lead to an improvement when accounting for shot noise and further explore behaviour when choosing different cost functions.

The manuscript is structured in the following way: First, we give a brief introduction to shadow tomography (Sec .~\ref {sec:shadows}) and v2rdm (Sec .~\ref {sec:v2rdm}). The former is focused on fermionic tomography respecting particle number and spin-preservation, and the latter briefly introduces core concepts of the constraints respected in our optimization and how to deal with inequalities when casting semidefinite programs. We then describe the two additional constraints derived from quantum data (Sec.~\ref{sec:shadowconstraints}), which we consider in our numerics, one from previous work~\cite{avdic23:_fewer_n} and one introduced in this work. We then simulate performance for various systems (Sec.~\ref{sec:numerics}) and compare it with the pure classical estimate and the pure quantum estimate derived from shadow tomography without re-optimising the 2-RDM to respect the N-representability conditions.

\section{Classical shadow protocol}\label{subsubsec:low_shadows}
\label{sec:shadows}
Classical shadows, a scheme introduced in~\cite{Huang_2020} based on the shadow tomography proposal~\cite{aaronson17:_shadow_tomog_quant_states}, can predict many properties of a quantum system with a number of measurements that scale logarithmically with the number of properties measured. 
The task of the protocol is to estimate the expectation value of multiple observables $\langle O_i \rangle$ on an unknown quantum state $\rho$. This requires the preparation of multiple copies of the state measured under random unitaries $U$ drawn from an ensemble $\mathcal{U}$, resulting in the measurement outcome $\ket{b}$. Averaging over the results constitutes the measurement channel 
\begin{align}
        \mathcal{M} & (\rho) = \nonumber\\ &
        \mathds{E}_{U \sim \mathcal{U}} \left[
        \sum_{b \in \{0,1\}^N}
        \bra{b} U\rho U^\dagger \ket{b} U^\dagger \ketbra{b}{b} U~\right].
\end{align}
\label{eq:shadow_protocol_1}
The collection of random unitaries $U$ and associated measurement outcomes $\bra{b}$ is referred to as a classical shadow. For each of these pairs, the measurement channel can be inverted to build an unbiased estimator of the quantum state
\begin{equation}
        \hat{\rho} = \mathcal{M}^{-1} \left(U^\dagger |b\rangle\langle b| U \right).
\end{equation}
From this classical estimator, one can infer the expected values of the desired observables $\expval{O_i} = \Tr(O_i\rho)$. The quality of the inverse of the measurement channel $\mathcal{M}^{-1}$ and the efficiency of the classical and quantum procedures depend on the chosen ensemble of unitaries and the associated observables one estimates from the quantum state. 

Multiple choices for the ensembles and associated classical estimation schemes have been proposed~\cite{zhao20_fermion_partial_tomog_via_class_shadow, Wan_2022,low22_class_shadow_fermion_with_partic_number_symmet}. We focus here on shadow estimation schemes twirling over the particle number restricted ensemble of matchgates~\cite{low22_class_shadow_fermion_with_partic_number_symmet}. This constitutes an efficient estimation scheme independent of the number of qubits, $N$, and its quantum cost scales only in the number of occupied fermionic modes $\eta$, which is the particle number of the trial state. This ensemble is usually referred to as the ensemble of single-particle basis rotations or orbital rotations $U(u)$, which define all valid transformations between different sets of the fermionic creation and annihilation operators $a_q$ and $a^\dag_q$ by
\begin{equation}
 U^\dagger(u) a_p^{\dagger} U(u)=\sum_{q=1}^{N}u_{pq}a_q^{\dagger}~,
\end{equation}
where $u \in \mathbb{C}^{N \times N}$ is a unitary matrix defining the change of basis of the fermionic modes, with their representation $U(u) \in \mathbb{C}^{2^N \times 2^N}$ in Hilbert space. We refer to the submatrix restricted to the $k$-particle subspace as $U_k \in \mathbb{C}^{\binom{N}{k}\times \binom{N}{k}}$, and therefore $U_{k=2}$ as the restriction to the two-particle subspace. We further restrict unitaries beyond preserving particle number to also preserve spin in Appendix \ref{app:spin_restr_shadows} as these are the most important symmetries for molecular systems.



In general, expectation values of generic $k$-particle reduced density matrices ($k$-RDM) can be inferred from this procedure~\cite{low22_class_shadow_fermion_with_partic_number_symmet}, we will only handle the two($k=2$)-particle reduced density matrices as no higher order density matrices are required for the purpose of estimating the observables of our interest. Our task, therefore, is to estimate the quantity:

\begin{equation}
\left\langle^2_S\mathbf{\hat{D}}^{pqrs}\right\rangle=\operatorname{Tr}\left[a^\dag_p a^\dag_q a_s a_r\rho\right],
\end{equation}
for any quantum state $\rho$. $^2_S\mathbf{\hat{D}}^{pqrs}$ is the shadow estimator of the 2-RDM, and the subscript prefix $S$ differentiates it from, e.g., the analytic 2-RDM $^2\mathbf{D}^{pqrs}$  of a quantum state in the following.

The single-shot estimator for the expectation value of the $2$-RDM for a measurement string $b$ in the computational basis under the basis transformation generated by $u$ is given by~\cite{low22_class_shadow_fermion_with_partic_number_symmet}


\begin{equation}
\label{eq:lowsingleshotestimator}
^2_S\mathbf{\hat{D}}^{pqrs}=\left\langle r,s\left|U_{k=2}\left(v_b u\right) E_{\eta, {k=2}} U_{k=2}^{\dagger}\left(v_b u\right)\right| p,q\right\rangle,
\end{equation}

where the basis states $ \bra{r,s}$ and $\ket{p,q}$ are defined by $\bra{r,s} =  \bra{0} a_{s}a_{r}$ and $\ket{p,q} =   a_{p}^{\dagger}  a_{q}^{\dagger}  \ket{0} $, indexing elements of the reduced density matrix. The matrix $v_b$ implements a rotation of $b$ into the ordered string $\{11\cdots 00\} \:= \{1^{\otimes \eta}\}+ \{0^{\otimes (N-\eta)}\}$ and the operator $E_{\eta, k}$ is defined as
\begin{equation}
 E_{\eta, k=2} = \sum_{p,q \in \{1,\ldots,N\}}|p,q\rangle\langle p,q| \frac{\left(\begin{array}{c}
\eta-s^{\prime} \\
2-s^{\prime}
\end{array}\right)\left(\begin{array}{c}
N-\eta+s^{\prime} \\
s^{\prime}
\end{array}\right)}{(-1)^{2+s^{\prime}}\left(\begin{array}{c}
k \\
s^{\prime}
\end{array}\right)} .
\end{equation}
with $s^\prime(p,q)$ being the number of shared indices with the reference determinant / Hartree Fock $|p=1,q=1\rangle$.
$$
s^\prime(p,q) = 
\begin{cases} 
      2 & p = 1 \wedge \, q = 1  \\
      1 &  p = 1 \vee \, q = 1  \\
      0 &  p \neq 1 \wedge \, q \neq 1 \\
   \end{cases}
$$
The \textit{average} shot variance of the 2-RDM estimator in equation ~\ref{eq:lowsingleshotestimator} is then bounded by
\begin{align}
\label{eq:lowbound}
&\mathbb{E}_{pqrs}\left[\operatorname{Var}\left[\left\langle^2_S\mathbf{\hat{D}}^{pqrs}\right\rangle\right]\right] \nonumber\\
&\quad \quad\leq\left(\begin{array}{l}
\eta \\
2
\end{array}\right)\left(1-\frac{\eta-2}{N}\right)^2\left(\frac{1+N}{1+N-2}\right),
\end{align}
which scales as $\mathcal{O}(\eta^2)$. The estimator in \ref{eq:lowsingleshotestimator} is only classically efficient for constant $k$ \cite{low22_class_shadow_fermion_with_partic_number_symmet}, as is the case here for $k=2$.

\section{v2rdm}
\label{sec:v2rdm}
Variational methods using the 2-RDM as a more compact stand-in for the wave function, as it contains all the information to calculate the energy, molecular forces, dipole moments, etc., were developed at the beginning of the 1950s~\cite{L_wdin_1955}. Originally, obtaining energies that were too low~\cite{Mayer_1955}, as unrestricted optimization results in 2-RDMs not stemming from a physical wave function, was violating the variational principle. Once embued with the concept of \textit{N-representability} conditions, restricting the space of obtainable 1- and 2-RDMs in optimization, conditions in an ever-increasing order were developed to guide variational methods to obtain physical results when traversing the space of 2-RDMs, by e.g., a semidefinite program \cite{Nakata_2001}. Recent high-performance implementations using GPUs and CPUs of these methods have led to the capability to treat larger-scale systems, e.g., 3,7-circumjacent in a (64e, 64o) active space~\cite{iii23:_variat_deter_two_elect_reduc_densit_matrix}. A brief treatment of the theory behind the workings of the algorithm will follow, but there are further resources~\cite{iii23:_variat_deter_two_elect_reduc_densit_matrix} for a more in-depth introduction to the topic.

In general, v2rdm is based on a convex optimization in the form of a semidefinite program. The merit/cost function is formulated as a vector $\mathbf{c}$, containing the coefficients to score a solution vector $\mathbf{x}$ over an inner product $\mathbf{c}^T \mathbf{x}$ to be minimized. We use the energy under the Hamiltonian $E\left[{ }^2 \mathbf{D}\right]$ as the cost function. We compare different choices of cost functions in Appendix \ref{app:cost}. The N-representability conditions are formulated as a system of linear equations, containing the set of linear equations embedded into the rows of a matrix $\mathbf{A}$ and their respective solutions as a solution vector $\mathbf{b}$, while the entries of $\mathbf{x}$ are cast into a block diagonal form in a matrix $M(\mathbf{x})$, where the matrix and therefore the individual blocks are to be kept positive semidefinite during optimization. The full set of equations is summarized as

\begin{equation}
    \begin{array}{rll}
\min _{\mathbf{x}} & \mathbf{c}^T \mathbf{x} & \\
\text { such that } & \mathbf{A x} & =\mathbf{b} \\
\text { and } M(\mathbf{x}) & \succeq 0.
\end{array}
\end{equation}

In our case, the individual elements of $x$ are associated with the individual entries of the 1-RDM $^1\mathbf{D}$ and 2-RDM $^2\mathbf{D}$, and further derived RDM-like quantities like the hole-particle 1-RDM $^1\mathbf{Q}^{pq} = \langle\psi|a_p a^\dag_q |\psi\rangle$ and similar quantities derived from the 2-RDM, that appear in more elaborate N-representability conditions \cite{avdic23:_fewer_n}. These are often referred to as the Q- and G-conditions, which come at an increased computational cost but lead to higher precision in the end result. Further constraints using the 3-RDM entries have been developed but won't be covered here. The block diagonal structure defined by $M(\mathbf{x})$ takes the form of

\begin{equation}
M(\mathbf{x})=\left(\begin{array}{cccccc}
{ }^1 \mathbf{D} & 0 & 0 & 0 & 0 & 0 \\
0 & { }^1 \mathbf{Q} & 0 & 0 & 0 & 0 \\
0 & 0 & { }^2 \mathbf{D} & 0 & 0 & 0 \\
0 & 0 & 0 & { }^2 \mathbf{Q} & 0 & 0 \\
0 & 0 & 0 & 0 & { }^2 \mathbf{G} & 0 \\
0 & 0 & 0 & 0 & 0 & \ldots
\end{array}\right) \succeq 0
\end{equation}

As there is an exponential amount \cite{prince:_libsdp} of N-representability conditions, enforcing all of them is computationally unfeasible, and one usually truncates the conditions at conditions of second or third order. At the second order, there are efficient algorithms in solving the resulting semidefinite program scaling as $\mathcal{O}(N^4)$ in memory and $\mathcal{O}(N^6)$ in compute, e.g. a boundary-point method ~\textit{BDSDP}~\cite{Povh_2006,Malick_2009,Mazziotti_2011}, and a matrix-factorization based procedure \textit{RRSDP}~\cite{Mazziotti_2004}. In total, the N-representability conditions usually considered in v2rdm can be summarized in the following way \cite{avdic23:_fewer_n}

\begin{align}
\min_{{ }^2 \mathbf{D}} & \, E\left[{ }^2 \mathbf{D}\right] \\
\text{such} \, \text{that} \quad { }^2 \mathbf{D} & \succeq 0 \\
{ }^2 \mathbf{Q} & \succeq 0 \\
{ }^2 \mathbf{G} & \succeq 0 \\
\operatorname{Tr}\left({ }^2 \mathbf{D}\right) & =\eta(\eta-1) \\
{ }^2 \mathbf{Q} & =f_Q\left({ }^2 \mathbf{D}\right) \\
{ }^2 \mathbf{G} & =f_G\left({ }^2 \mathbf{D}\right).
\end{align}

$f_G$ and $f_Q$ are linear mappings between $^2 \mathbf{G}$,$^2 \mathbf{Q}$ and 
$^2 \mathbf{D}$. We use a spin-adapted v2rdm code and also enforce spin-adapted constraints from the classical shadow; for further details when adapting for spin, consult the Appendix \ref{app:spin_restr_v2rdm}. To these traditional constraints, we add now additional constraints stemming from the classical shadow measurements on a quantum computer.

\section{Shadow constraints}
\label{sec:shadowconstraints}
We consider two different sets of constraints stemming from the output of a quantum computer, further constraining the optimization problem. The first way of constraining the optimization problem relies on the classical shadow estimator in Eq. \ref{eq:lowsingleshotestimator}, that bounds the variance of the average 2-RDM element in Eq. \ref{eq:lowbound}. As the variance properties of the classical shadow protocol rely on using the estimator this is optimal in terms of quantum measurements. We refer to this constraint to as constraint (1) from here. We build an $\epsilon$-ball around all the elements of the true 2-RDM $^2\mathbf{D}^{pqrs}$ from the estimator $^2_S\mathbf{\hat{D}}^{pqrs}$ in equation ~\ref{eq:lowsingleshotestimator} and estimating the radius of the ball $\epsilon_1$ by means of the upper bounds of the variance from equation~\ref{eq:lowbound}.

\begin{equation}
    ^2_S\mathbf{\hat{D}}^{pqrs}-\epsilon_1 \leq \,^2\mathbf{D}^{pqrs} \leq \,^2_S\mathbf{\hat{D}}^{pqrs} + \epsilon_1
    \label{eq:bi_constr}
\end{equation}

This results in $2N^4$ additional constraints to the optimization, which are of constant size. These can further be divided into restrictions on each spin sector as discussed in Appendix \ref{app:spin_restr_v2rdm}. In general, the classical resources needed to calculate the estimator stemming from the classical shadow in equation ~\ref{eq:lowsingleshotestimator} scale as $\mathcal{O}(N^5)$ in the classical post-processing compared to the scaling $\mathcal{O}(N^6)$ of the semidefinite program. The classical optimization, therefore, adds a linear multiplicative overhead in classical computational cost while potentially reducing the number of quantum resources/shots required to reach a given precision. There exists a more efficient way of estimating from a classical shadow using Pfaffians ~\cite{Wan_2022,low22_class_shadow_fermion_with_partic_number_symmet} not described here, which could bring this cost down even more.

The second constraint, referred to as constraint (2), introduced in~\cite{avdic23:_fewer_n} is based on the diagonal part of the 2-RDM, measured in the computational basis under $m$ random single-particle rotations $U_m$ from~\cite{low22_class_shadow_fermion_with_partic_number_symmet}

\begin{equation}
S_m^{p q}=\left\langle\Psi\left|\hat{U}_m^{\dagger} \hat{a}_p^{\dagger} \hat{a}_q^{\dagger} \hat{a}_q \hat{a}_p \hat{U}_m\right| \Psi\right\rangle.
\end{equation}

This constraint also relies on randomized measurements of the quantum state similar to (1) but not using any of the improved estimators provided by the classical shadow protocol. This leads to $2mN^2$ additional constraints, with $N^2$ entries. The factor of 2 follows from constraining both sides of the inequality (see App. \ref{app:ineq_sdp} for more details):

\begin{equation}
S_m^{p q}-\epsilon_2 \leq X_m^{p q} \leq S_m^{p q}+\epsilon_2,
\label{eq:mazz_constr}
\end{equation}
with
\begin{equation}
X_m^{p q}=\left(\left(U_m \otimes U_m\right)\, ^2D\left(U_m \otimes U_m\right)^T\right)^{p q},
\end{equation}

 where $\epsilon_2$ marks the error on each element due to shot noise and device error. We only investigate behaviour under shot noise considerations in this manuscript, but as device error can also be modelled under similar Gaussian perturbations to the correct RDM, one could further study behaviour under noisy conditions ~\cite{avdic23:_fewer_n,Rubin_2018}.

\section{Numerics}
\label{sec:numerics}
In this section, we numerically investigate the performance of the proposed optimization of the 2-RDM when using the two different constraints (equation ~\ref{eq:bi_constr} and equation ~\ref{eq:mazz_constr}, labelled (1) and (2))  stemming from the classical shadow estimation in comparison with the estimate stemming from a vanilla v2rdm estimation without any additional constraints and an estimate stemming from a classical shadow estimation without any v2rdm post-optimization of the estimate (equation ~\ref{eq:lowsingleshotestimator}). To allow for comparison between the two constraints, their measurement shot budgets are considered in the following way: For more efficient numerical considerations, the estimate from a classical shadow for (1) is simulated by adding Gaussian noise to the 2-RDM from an full configuration interaction (FCI) calculation where the strength of the perturbation on each element is calculated by the bound of the shot variance of each element from equation ~\ref{eq:lowbound} converted into a standard deviation by dividing by the number of shots taken at the respective point. Previous work~\cite{scheurer23:_tailor_exter_correc_coupl_clust_quant_input,kiser23:_class_quant_monte_carlo} has shown this to be an accurate noise model for the estimate of the classical shadow.  As the constraint for (2) in equation ~\ref{eq:mazz_constr} relies on measuring the diagonal elements of the 2-RDM under random basis rotations, we consider the average variance of one of these diagonal elements. For e.g. the $^2D_{\alpha\beta}$ the expectation value of a diagonal element of the 2-RDM $\mathbb{E}[S_{pq}]$ can be calculated by
\begin{equation}
    \mathbb{E}[S^{pq}] = (N_\alpha N_\beta)^2/\eta^2, \quad \text{Var}[S^{pq}] = 1 - \mathbb{E}(S^{pq})^2,
\end{equation}
where $N_\alpha,N_\beta$ are the number of $\alpha$,$\beta$ electrons and $\eta=N_\alpha+N_\beta$. The respective variance $\text{Var}[S_{pq}]$ follows from the observable mapping down to a Pauli observable  ($Z$) when measuring in the computational basis after the random basis transformation. To allow for equal shots for both ways of constraining the problem, the number of basis $m$ sampled when constraining with equation ~\ref{eq:mazz_constr} is taken to fulfill
\begin{equation}
    \mathbb{E}_{pqrs}\left[\operatorname{Var}\left[\left\langle{}^2_S\mathbf{\hat{D}}^{pqrs}\right\rangle\right]\right] \approx m \text{Var}[S^{pq}]
\end{equation}
 rounded up to the closest integer $m$. As the bound of the shot variance given in equation ~\ref{eq:lowbound} is also only bounding the \textit{average} shot variance of an entry of the 2-RDM we deem this a fair comparison.
 
 One way to view the two different constraints is as follows: measuring the diagonal elements, one \textit{oversamples} measurements under the same basis rotation, so constraint (1) can be viewed as the limit of (2) when sampling in a new basis for every new shot and using the estimator in equation ~\ref{eq:lowsingleshotestimator}. As changing the basis requires reconfiguring the quantum circuit, this is usually associated with increased cost, and \textit{oversampling} has been explored in Ref.~\cite{scheurer23:_tailor_exter_correc_coupl_clust_quant_input} for potentially reducing cost in shadow estimation when accounting for this cost. We do not include this potentially increased cost here, but it can also be accounted for when constraining in (1) by increasing the bound of the variance from equation ~\ref{eq:lowbound}. Covariances are not considered here, as previous work has shown numerical evidence that no covariances are present when estimating orthogonal elements in classical shadow estimation using matchgates~\cite{scheurer23:_tailor_exter_correc_coupl_clust_quant_input,kiser23:_class_quant_monte_carlo}.

 Our implementation extends an open source version of a v2rdm code using libsdp, providing efficient solvers for the semidefinite program~\cite{iii23:_variat_deter_two_elect_reduc_densit_matrix,prince:_libsdp} with the constraints mentioned above. We further use Openfermion~\cite{mcclean17:_openf} to build the correct Hilbert space representation of our basis rotations in the correct subspace to build our constraints and simulate measurements on a quantum computer. Chemistry Hamiltonians and active space calculation are done by PySCF ~\cite{sun17:_python_simul_chemis_framew_pyscf}.

 We use our test systems, \ce{N2} in cc-pvdz basis in an (10,8) active space and benzene in the STO-3G in a (16,16) active space to investigate how system size affects the estimation. As previously mentioned, we are mostly concerned with improving the estimate of the 2-RDM from the shadow estimation scheme. Improvements in energy estimation we deem less interesting as one has access to more efficient ways of measuring the energy, e.g., double factorization~\cite{von_Burg_2021,oumarou22:_accel_quant_comput_chemis_throug} or tensor-hypercontraction~\cite{Lee_2021}. Constructing the Hilbert-Space unitaries needed for (2) becomes fairly costly at large system sizes, and consequently, we only investigate constraint (1) for the largest of the test systems, benzene.

\begin{figure}
    \centering
    \includegraphics[width=.4\textwidth]{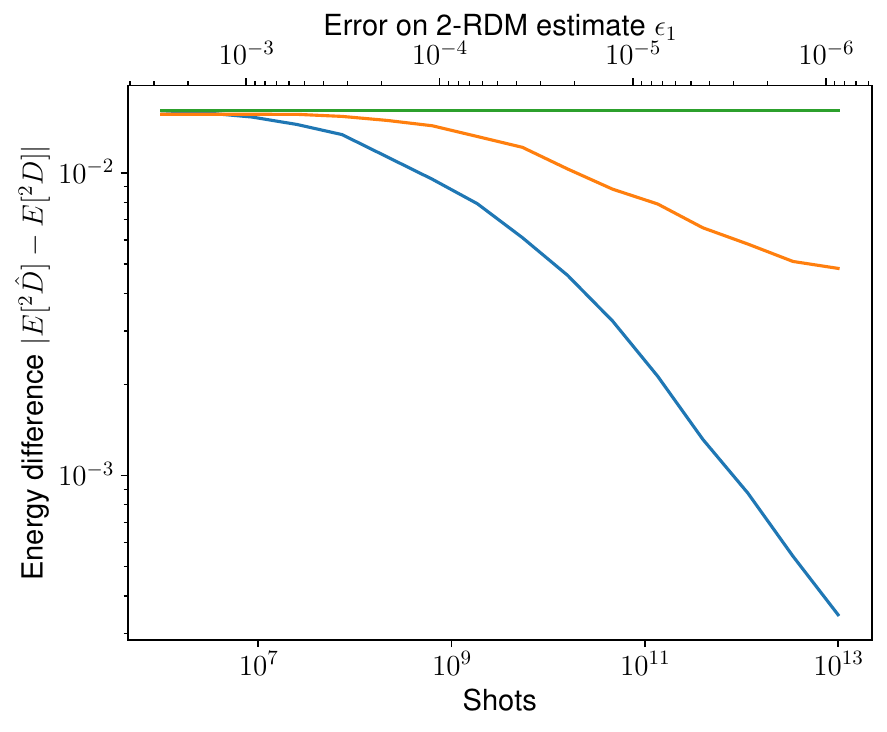}
    \includegraphics[width=.4\textwidth]{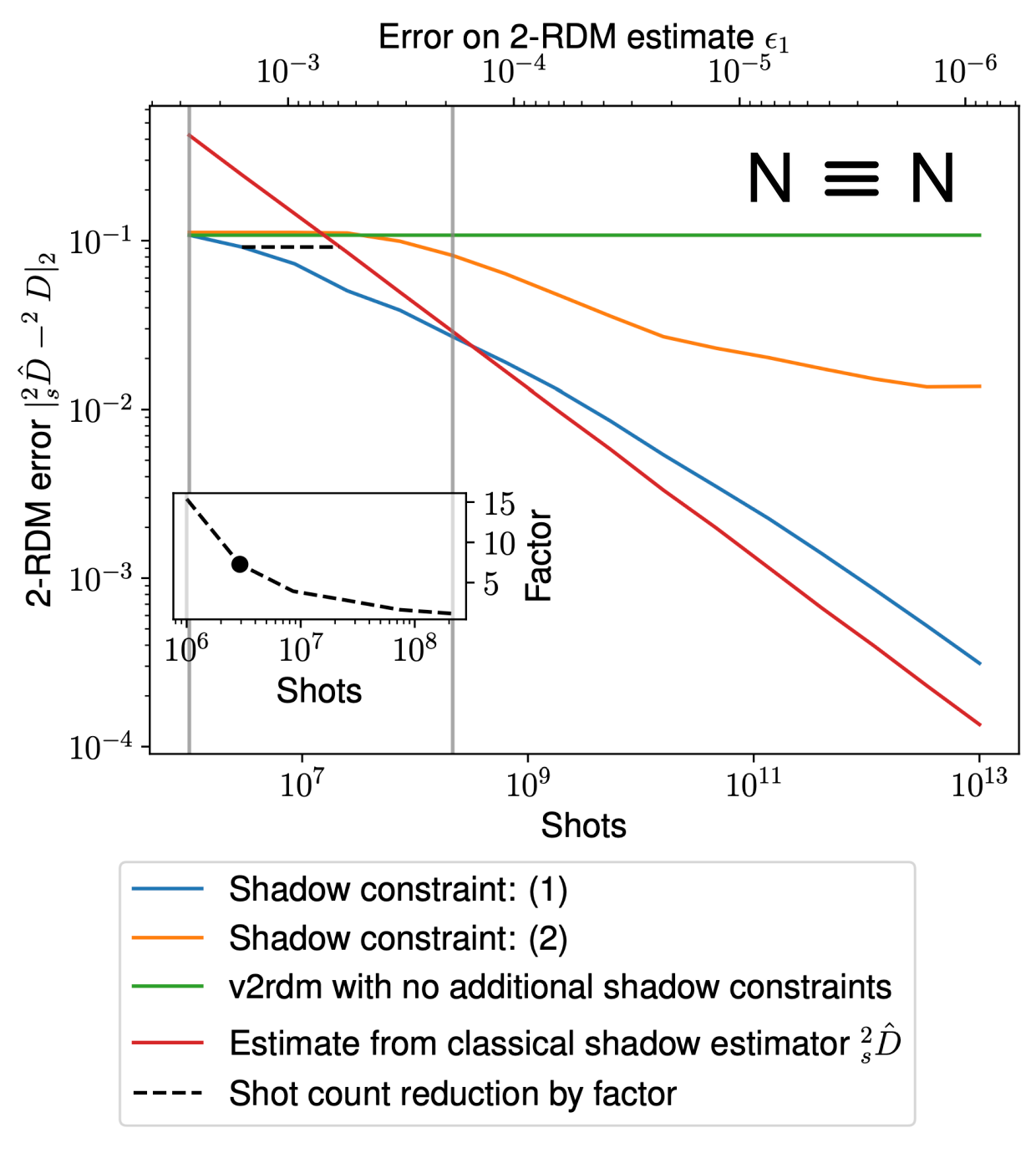}
    \caption{Top: Energy error of the estimate of the ground state energy for \ce{N2} in dependence on the number of shots spent on the quantum shadow constraints included in the optimization. Bottom: 2-RDM error in the Frobenius norm towards the true 2-RDM stemming from an FCI calculation of N2 in dependence on the number of shots  spent on the quantum shadow constraints included in the optimization. Inset quantifies the factor of improvement over the estimate coming just from the classical shadow at the same level of error (red). The dot in the inset corresponds to the dashed line in the outer figure as a guide to the eye and clarification.}
    \label{fig:n2}
\end{figure}

For the smaller active space (10,8) for \ce{N2}, looking at the energy, we find shadow constraint (1) outperforming shadow constraint (2) over the range of the shot interval investigated (Figure ~\ref{fig:n2}, top). For the estimate of the 2-RDM compared to the true 2-RDM (Figure ~\ref{fig:n2}, bottom) we also find shadow constraint (1) outperforming constraint (2) consistently. Above $10^6$ shots, shadow constraint (1) improves over the plain v2rdm implementation, whereas the unadapted shadow estimation only improves above $10^7$ over the purely classical estimate. Constraint (1) improves over the shadow estimate up until $2 \times 10^8$ shots, where the shadow estimator overtakes the combined estimate, reaching a reduction of up to a factor of 15 required shots to reach the same precision. The threshold at  $2\times 10^8$  can be explained by the energy acting as the cost function in optimization: At some point, the optimiser biases the estimate by improving the entries of the 2-RDM with high weight from the Hamiltonian at the cost of the other entries, losing precision over the unbiased estimate of the classical shadow estimation performed on the true ground state (see App. \ref{app:bias}). Shadow constraint (2) does not improve over v2rdm without constraints in the low shot regime or the estimate from classical shadow estimation at the $10^7$ shots threshold.

In Appendix \ref{app:energy_constr} adding more information from a fault tolerant calculation e.g. the estimate of the ground state energy is investigated.

\begin{figure}
    \centering
    \includegraphics[width=.4\textwidth]{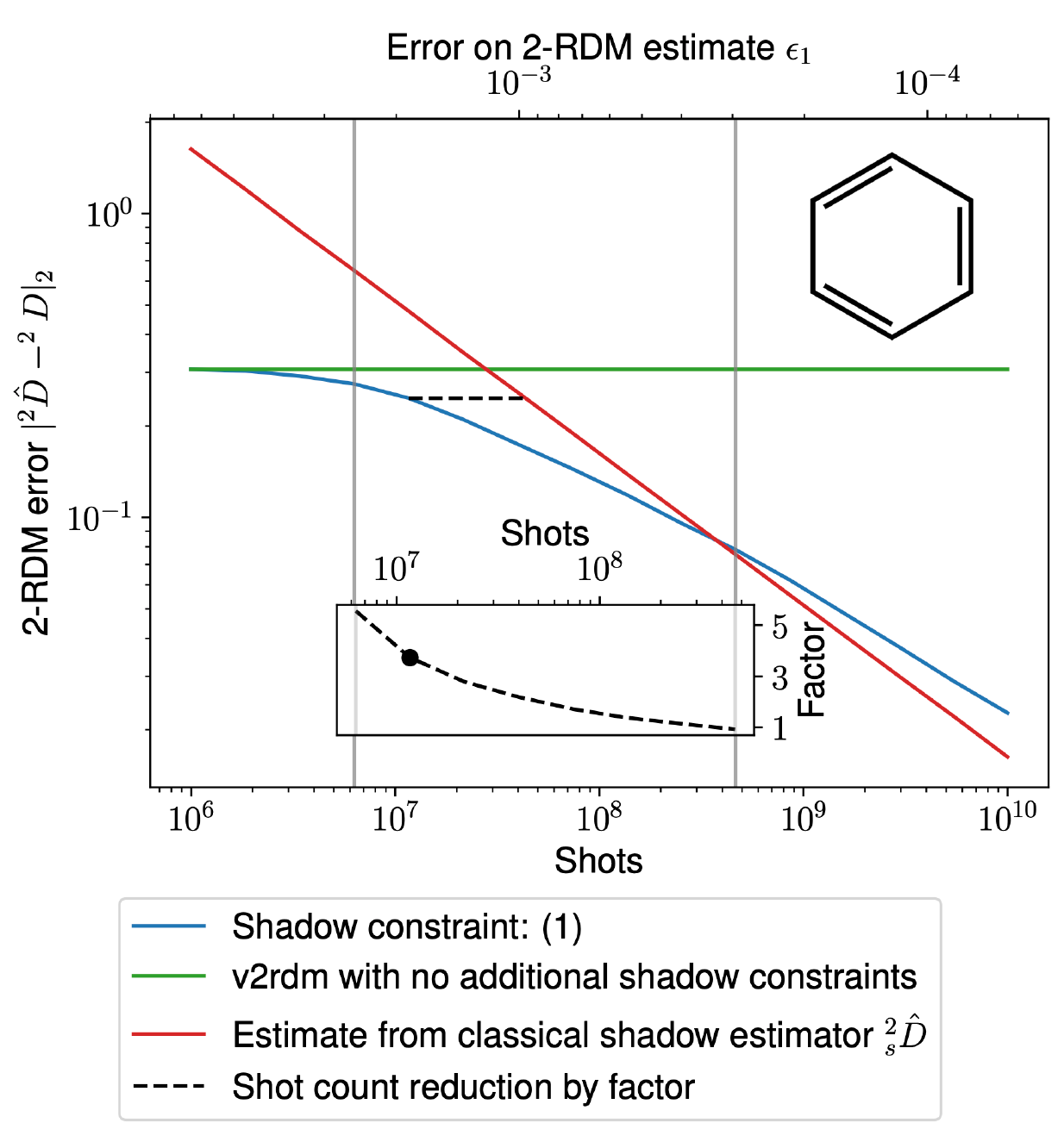}
    \caption{2-RDM error in the Frobenius norm towards the true 2-RDM stemming from an FCI calculation of benzene in a (16,16) active space in dependence of the number of shots spent on the quantum shadow constraints included in the optimization. Insert quantifies the factor of improvement over the estimate coming just from the classical shadow at the same level of error (red). The dot in the inset corresponds to the dashed line in the outer figure as a guide to the eye and clarification.}
    \label{fig:benzene}
\end{figure}

For the large active space (16,16) for benzene (Figure ~\ref{fig:benzene}), we find a qualitatively very similar looking relationship between plain v2rdm, classical shadow estimation, and v2rdm with shadow constraint (1): At an error that corresponds roughly with $10^{-3}$ the shadow estimation beats the purely classical estimate. v2rdm with shadow constraint (1) improves over both the other estimates for roughly two orders of magnitude of shot budgets before the bias introduced into the estimate loses over the unbiased shadow estimation technique. Improvement factors also seem to be a bit lower in this scenario, only reaching up to a factor of 5.

\section{Discussion}

To conclude, our simulations show that v2rdm can help classical shadow estimation to increase precision for a given shot budget in certain regimes. When accounting for shot noise, we find that constraint (2) of previous work \cite{avdic23:_fewer_n} does not improve the 2-RDM estimate over either the only classical optimization or quantum only estimator at a given shot budget. Only when constraining the optimization with the improved  constraint (1), we find up to a factor of 15 in improvement for \ce{N2}, although the factor of improvement is largest when being close to the result of the purely classical estimate. This improvement seems to be reduced for larger systems. For high precision estimates, the bias introduced seems to lose out over the unbiased estimate from shadow estimation. Nonetheless, when quantum resources are scarce in near- and mid-term shadow estimations trying to extract 2-RDM quantities from a state to extract, e.g., forces and other molecular properties beyond ground state energies, classically optimising the estimate from a classical shadow can lead to increased precision at a fixed quantum cost. Although the window where v2rdm actually improves the estimate is limited, for estimations where high precision is necessary, e.g., chemistry applications, this does not solve the cost of extracting a 2-RDM from a quantum device. As we have investigated scaling of epsilon under the standard quantum limit $\frac{1}{\epsilon^2}$, the cost savings would become much more dramatic for recently proposed procedures~\cite{king24:_tripl} with better scaling in $N$ at the cost of increased scaling in the error $\epsilon$.

As the optimization under the energy as a cost function seems to introduce a bias limiting the applicability window where the combined methods improve over pure shadow estimation, changing the cost function to, e.g., the 1-norm of the 2-RDM~\cite{Rubin_2018,avdic23:_fewer_n} might be an avenue in further improvement of the scheme presented here.

\section{Acknowledgments}
We thank Nicholas Rubin and Ryan Babbush for discussions on fermionic marginal constraints. Furthermore, we thank Christian Gogolin and Maximilian Scheurer for input on shadow estimation. We thank Eugene DePrince III and Brecht Verstichel for discussions about the v2rdm method.
\bibliography{lib}

\appendix

\section{Spin-restricted v2rdm}
\label{app:spin_restr_v2rdm}
As we work in a spin-adapted setting, one can further find a sub-block structure of the positive-semidefinite matrices. We decompose the 1-particle density matrix  $^1\mathbf{D}^{pq} = \langle\psi|a^\dag_p a_q |\psi\rangle$ and its derivate $^1\mathbf{Q}^{pq} = \langle\psi|a_p a^\dag_q |\psi\rangle$ into spin-blocks in the following way, as discussed in further depth in~\cite{Rubin_2018}
\begin{equation}
{ }^1 \mathbf{D}=\left(\begin{array}{cccccc}
{ }^1 \mathbf{D}_{\alpha} & 0   \\
0 & { }^1 \mathbf{D}_{\beta}  \\
\end{array}\right)
\end{equation}
and $Q$
\begin{equation}
{ }^1 \mathbf{Q}=\left(\begin{array}{cccccc}
{ }^1 \mathbf{Q}_{\alpha} & 0   \\
0 & { }^1 \mathbf{Q}_{\beta}.  \\
\end{array}\right)
\end{equation}
The two particle conditions of $^2\mathbf{D}^{pqrs} = \langle\psi|a^\dag_p a^\dag_q a_s a_r |\psi\rangle$ and derived quantities $^2\mathbf{Q}^{pqrs} = \langle\psi|a^\dag_p a_q a^\dag_s a_r |\psi\rangle$ and $^2\mathbf{Q}^{pqrs} = \langle\psi|a_p a_q a^\dag_s a^\dag_r |\psi\rangle$ can also further be block diagonalized into spin sectors. For compactness sake, we leave out the upper indices on the diagonal ${ }^2 \mathbf{D}_{\alpha\alpha} \equiv { }^2 \mathbf{D}_{\alpha\alpha}^{\alpha\alpha}$ and get four spin-blocks for the 2-RDM
\begin{equation}
{ }^2 \mathbf{D}=\left(\begin{array}{cccc}
{ }^2 \mathbf{D}_{\alpha\alpha} & 0 & 0 & 0   \\
0 & { }^1 \mathbf{D}_{\beta\beta} & 0 & 0 \\
0 & 0 & { }^2 \mathbf{D}_{\alpha\beta} & 0  \\
0 & 0 & 0 & { }^2 \mathbf{D}_{\beta\alpha}  \\
\end{array}\right)
\end{equation}
and a slightly more elaborate structure of G
\begin{equation}
{ }^2 \mathbf{G}=\left(\begin{array}{cccc}
{ }^2 \mathbf{G}_{ \alpha, \alpha} & { }^2 \mathbf{G}_{\alpha, \alpha}^{\beta, \beta} & 0 & 0 \\
{ }^2 \mathbf{G}_{\beta, \beta}^{\alpha, \alpha} & { }^2 \mathbf{G}_{\beta, \beta} & 0 & 0 \\
0 & 0 & { }^2 \mathbf{G}_{\alpha, \beta} & 0 \\
0 & 0 & 0 & { }^2 \mathbf{G}_{\beta, \alpha}
\end{array}\right)
\end{equation}
whereas Q has the same structure as G.


\section{Inequalities in semidefinite programming}
\label{app:ineq_sdp}
As we are constraining our semidefinite program now with estimates associated with a certain error $\epsilon$ coming from the statistical nature of the estimate, we have to be able to handle inequalities when constraining the optimization problem. To accommodate inequalities one converts these inequalities into equalities by using a set of ancillary variables $\varphi_0, \hdots \varphi_i$ that are added to the optimization in the form of individual 1 by 1 blocks.

\begin{equation}
M(\mathbf{x})=\left(\begin{array}{ccccccc}
{ }^1 \mathbf{D} & 0  & \hdots &  &  &  &  \\
0 & \ddots & \ddots & &  &  & \\
\vdots & \ddots & { }^2 \mathbf{D} & \ddots &  &  & \\
 &  & \ddots & \ddots & \ddots & & \\
 &  &  & \ddots & \varphi_0 & \ddots & \vdots \\
 &  &  &  & \ddots & \ddots & 0  \\
 &  &  &  & \hdots  & 0 & \varphi_i 
\end{array}\right) \succeq 0
\end{equation}

The $j$th inequality from equation ~\ref{eq:bi_constr} constraining the problem

\begin{align}
    ^2_S\mathbf{\hat{D}}^{pqrs} -\epsilon &\leq \,^2\mathbf{D}^{pqrs} \leq \,^2_S\mathbf{\hat{D}}^{pqrs} + \epsilon,
    \label{eq:bi_constr_spin}   
\end{align}

can then be converted into two equalities by adding an ancillary variable to a side of each inequality to equalize both sides 

\begin{align}
    ^2_S\mathbf{\hat{D}}^{pqrs} -\epsilon &= \,^2\mathbf{D}^{pqrs} - \varphi_{2j}\\
    ^2\mathbf{D}^{pqrs} + \varphi_{2j+1} &= \,^2_S\mathbf{\hat{D}}^{pqrs}  + \epsilon,
\end{align}
where $j$ is an index keeping track of labelling individual sets of $p,q,r,s$, assigning it its own pair of ancillary variables. This holds as long as $\varphi_i >0$ which is guaranteed by keeping the individual ancillary variables $\varphi_i$ in an individual $(1x1)$ block of $M(x)$ during the optimization, and the maximum error on each element can be upper bounded by equation ~\ref{eq:lowbound} for constraint (1).

In the spin adapted way of constraining the semidefinite program, our condition from equation ~\ref{eq:bi_constr} becomes a restriction on the individual spin-full 2-RDMs

\begin{align}
    ^2_S\mathbf{\hat{D}}^{pqrs}_{\alpha\alpha} -\epsilon &\leq \,^2\mathbf{D}^{pqrs}_{\alpha\alpha} \leq \,^2_S\mathbf{\hat{D}}^{pqrs}_{\alpha\alpha}  + \epsilon, \label{eq:bi_constr_spin_alphaalpha}\\
     ^2_S\mathbf{\hat{D}}^{pqrs}_{\alpha\beta} -\epsilon &\leq \,^2\mathbf{D}^{pqrs}_{\alpha\beta} \leq \,^2_S\mathbf{\hat{D}}^{pqrs}_{\alpha\beta}  + \epsilon, \\
      ^2_S\mathbf{\hat{D}}^{pqrs}_{\beta\beta} -\epsilon &\leq \,^2\mathbf{D}^{pqrs}_{\beta\beta} \leq \,^2_S\mathbf{\hat{D}}^{pqrs}_{\beta\beta}  + \epsilon.
\end{align}

Again, each of these constraints can be converted into two equalities by adding an ancillary variable to a side of the inequality, here shown for one of the $(\alpha,\alpha)$ constraints \ref{eq:bi_constr_spin_alphaalpha}.

\begin{align}
    ^2_S\mathbf{\hat{D}}^{pqrs}_{\alpha\alpha} -\epsilon &= \,^2\mathbf{D}^{pqrs}_{\alpha\alpha} - \varphi_{2i,\alpha\alpha}\\
    ^2\mathbf{D}^{pqrs}_{\alpha\alpha} + \varphi_{2i+1,\alpha\alpha} &= \,^2_S\mathbf{\hat{D}}^{pqrs}_{\alpha\alpha}  + \epsilon.
\end{align}

More details on how to extract spin-full RDMs from the classical shadow is found in the next chapter \ref{app:spin_restr_shadows}.

\section{Spin-restriction of classical shadows}
\label{app:spin_restr_shadows}

One can further adapt the estimator in ~\ref{eq:lowsingleshotestimator} and the shadow protocol to respect another symmetry apart from particle number - spin. One can restrict the Haar-random unitary generating the basis rotation $u_{pq}$ into a block-diagonal form,~\cite{zhao23_group_theor_error_mitig_enabl}
\begin{equation}
u_{pq}=\left(\begin{array}{cc}
u_{pq,\alpha} & 0 \\
0 & u_{pq,\beta}
\end{array}\right),
\end{equation}
which allows for a more efficient way of implementing the unitary in terms of Givens rotations into two local Givens circuits reducing the circuit depth needed to implement the measurement rotation further by a factor of 2. As the classical post-processing time scales with the size of the Hilbert-subspace, one can carefully restrict the estimator to a smaller subspace

\begin{align}
 \notag E^{\alpha,\beta}_{\eta, 2} &= \\ \sum_{p,q \in \{1,\ldots,N_\text{orb}\}}&|p^\alpha,q^\beta\rangle\langle p^\alpha,q^\beta| \frac{\left(\begin{array}{c}
\eta-s^{\prime} \\
k-s^{\prime}
\end{array}\right)\left(\begin{array}{c}
N-\eta+s^{\prime} \\
s^{\prime}
\end{array}\right)}{(-1)^{k+s^{\prime}}\left(\begin{array}{c}
k \\
s^{\prime}
\end{array}\right)},
\end{align}
likewise for $(\alpha,\alpha)$, $(\beta,\alpha)$ and $(\beta,\beta)$ to only index the $\alpha/\beta$ part of the $p/q$ spatial orbital. This reduces the size of the individual subspaces by a factor of $2$ and the number of entries by factor of $4$. 

\section{Additional numerics}

\begin{figure}
    \centering
    \includegraphics[width=.4\textwidth]{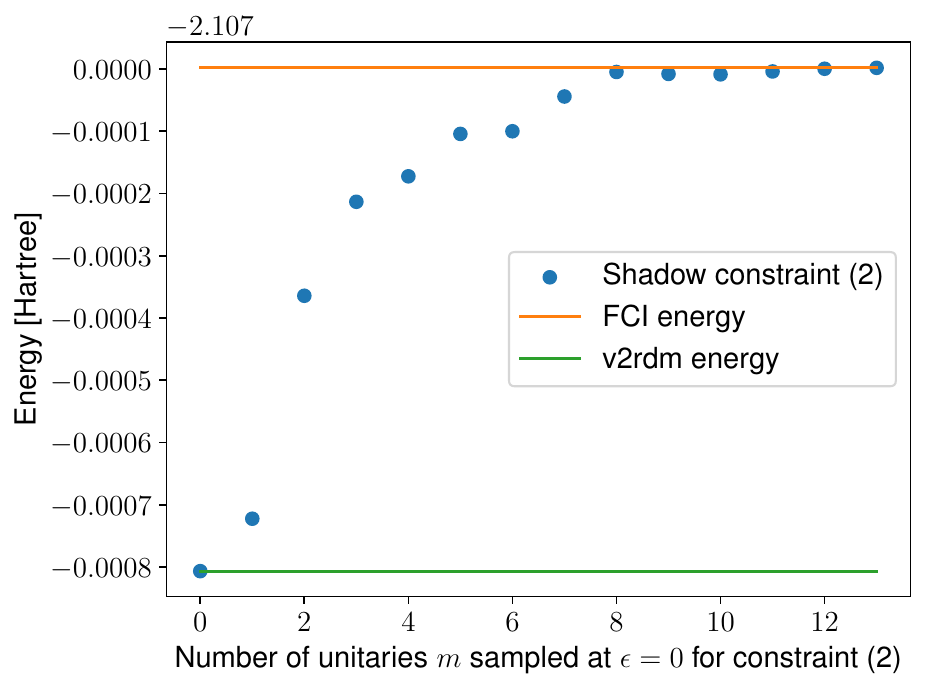}
    \caption{Energy convergence for $\text{H}_4$ for the optimization under shadow constraint (2) in dependence of number of basis $m$ used for additional constraints during optimissation.}
    \label{fig:h4}
\end{figure}

To verify our implementation of shadow constraint (2), we also run simulations in the scenario considered most of the time in the reference~\cite{avdic23:_fewer_n} at $\epsilon_2 = 0$. We study H4 in STO-3G the energy convergence in Figure \ref{fig:h4} in dependence of the number of basis sampled $m$, also reaching exact agreement at $m=12$ basis used for the additional constraints during optimization at infinite precision, corresponding to taking an infinite number of shots at each basis. So we deem our implementation of the shadow constraint comparable to the one studied in previous work.

\section{Introduced bias in optimization}
\label{app:bias}

\begin{figure}
    \centering
\includegraphics[width=.4\textwidth]{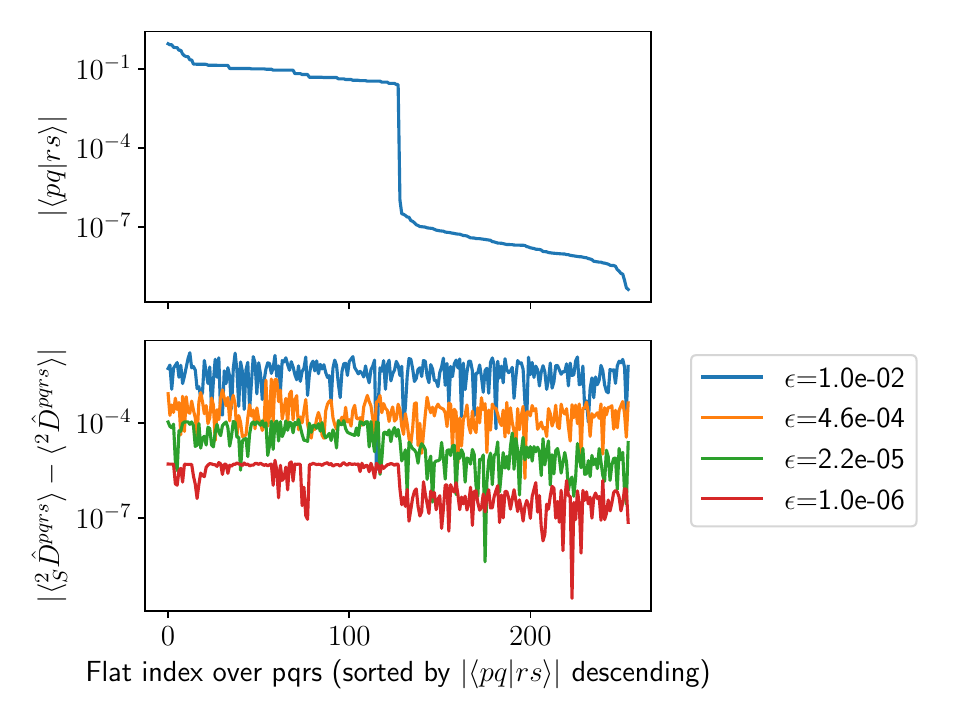}
    \caption{Top: Absolute value of the 2-body coefficients of the Hamiltonian  / two electron integrals for $\text{H}_4$  in STO-3G sorted descending. Bottom: Difference between the shadow estimate at a given epsilon used to build the shadow constraints and the 2-RDM estimate after the semidefinite optimization averaged over 5 runs sorted by the ordering of the upper plot.}
    \label{fig:bias}
\end{figure}

As our numerics show, when including a high precision estimate of a classical shadow with a tight $\epsilon$, the optimization actually increases the distance to the true 2-RDM during optimization (Fig. \ref{fig:h4}). Here we investigate the mechanism in how the bias is introduced. We look at the elements of the 2-RDM after the semidefinite optimization at different precisions of the ingoing shadow data. We linearize the matrix and sort them by the magnitude of the associated coefficients (see. Fig. \ref{fig:bias} top) in the Hamiltonian that constitutes the cost function used in the optimization.

We plot the difference of each element to the estimate from the classical shadow estimator used to generate the additional constraints. As the shadow data is generated by applying Gaussian noise to the correct 2-RDM, each element will be on average equally far away from its true value and therefore should be moved similarly during optimization. At low precision (Fig. \ref{fig:bias} bottom, blue) we can see that elements associated with big and small coefficients in the Hamiltonian get changed in the optimization on similar magnitudes, therefore no or at least a very small bias is introduced compared to the variance of the estimator. At high precision (Fig. \ref{fig:bias} bottom, red) one can clearly identify the point at which the Hamiltonian coefficients decrease by six orders of magnitude in the bottom plot. To decrease the energy which constitutes the cost function, terms with high weight get moved an order of magnitude more during the optimization than lower weight terms introducing a bias to the result, as in an unbiased optimization we expect each element to be moved similarly during optimization.

\section{Choice of cost function}
\label{app:cost}
When variationally optimizing the 2-RDM with regards to N-representability constraints, most implementations ~\cite{Mazziotti_2011, avdic23:_fewer_n, prince:_libsdp} minimize the energy $\min_{{ }^2 \mathbf{D}} \, E\left[{ }^2 \mathbf{D}\right]$ with regard to the Hamiltonian as a cost function ~\cite{Mazziotti_2011,prince:_libsdp}. Other forms of cost functions have been used very successfully in other applications of semi definite programming, e.g. when completing matrices with only partially known entries ~\cite{Candes_2008} and recovering matrices from noisy estimates of their entries ~\cite{peng23:_fermion}. This is based on the finding that low-rank approximations of these matrices are on average a good representation of the unknown matrix. As minimizing the rank is NP-hard \cite{Chi_2018}, the first relaxation to a convex manifold being the trace or nuclear norm $\|\cdot\|_*$ is used as a heuristic in optimization \cite{Chi_2018,Gross_2011}. The Frobenius norm $\|\cdot\|_F$ also serves as a possible cost function in these applications \cite{avdic23:_fewer_n}. For a (positive semidefinite) matrix $A$ the norms are defined as 

\begin{align}
    \|A\|_*&=\operatorname{Tr}\left(\sqrt{A^* A}\right)\\
    \|A\|_F&=\sqrt{\operatorname{Tr}\left(A^* A\right)}.
\end{align}

These type of cost functions have also been applied in quantum chemistry settings, as 2-RDMs of ground states are known to be approximately low rank ~\cite{Schwerdtfeger_2012,Gidofalvi_2007,peng23:_fermion} and have been investigated in improving readout when sampling in the Pauli basis ~\cite{peng23:_fermion}, finding that in the noisy readout setting it was optimal to sample all terms and the optimization mostly acting as a low rank filter to the estimation. As the shadow estimation used here estimates all Fermionic terms with equal error, optimizing for nuclear norm also seems as a reasonable choice as a cost function.

We investigate the following different cost functions in our setting 

\begin{align}
\min_{{ }^2 \mathbf{D}} & \, E\left[{ }^2 \mathbf{D}\right] \label{eq:nrg_cost} \\
\min_{{ }^2 \mathbf{D}}& \, E\left[{ }^2 \mathbf{D}\right] -   || ^2 \mathbf{D} ||_* \label{eq:nrg_nuc_cost}\\
\min_{{ }^2 \mathbf{D}} &  \, ||^2 \mathbf{D}||_* \label{eq:nuc_cost} \\
\min_{{ }^2 \mathbf{D}} &  \, ||^2 \mathbf{D}||_F \label{eq:fro_cost}
\end{align}

There are a total of four different cost functions we consider: 
\begin{enumerate}
    \item the energy under the Hamiltonian \ref{eq:nrg_cost}
    \item the energy under the Hamiltonian regularized by the nuclear norm (schatten 1-norm or trace norm) $||\cdot||_1$ of the 2-RDM, therefore penalizing solutions with higher nuclear norm \ref{eq:nrg_nuc_cost}
    \item the nuclear norm $||\cdot||_1$ of the 2-RDM \ref{eq:nuc_cost}
    \item the Frobenius norm $||\cdot||_F$ of the 2-RDM \ref{eq:fro_cost}
\end{enumerate}

\begin{figure}
    \centering
\includegraphics[width=.4\textwidth]{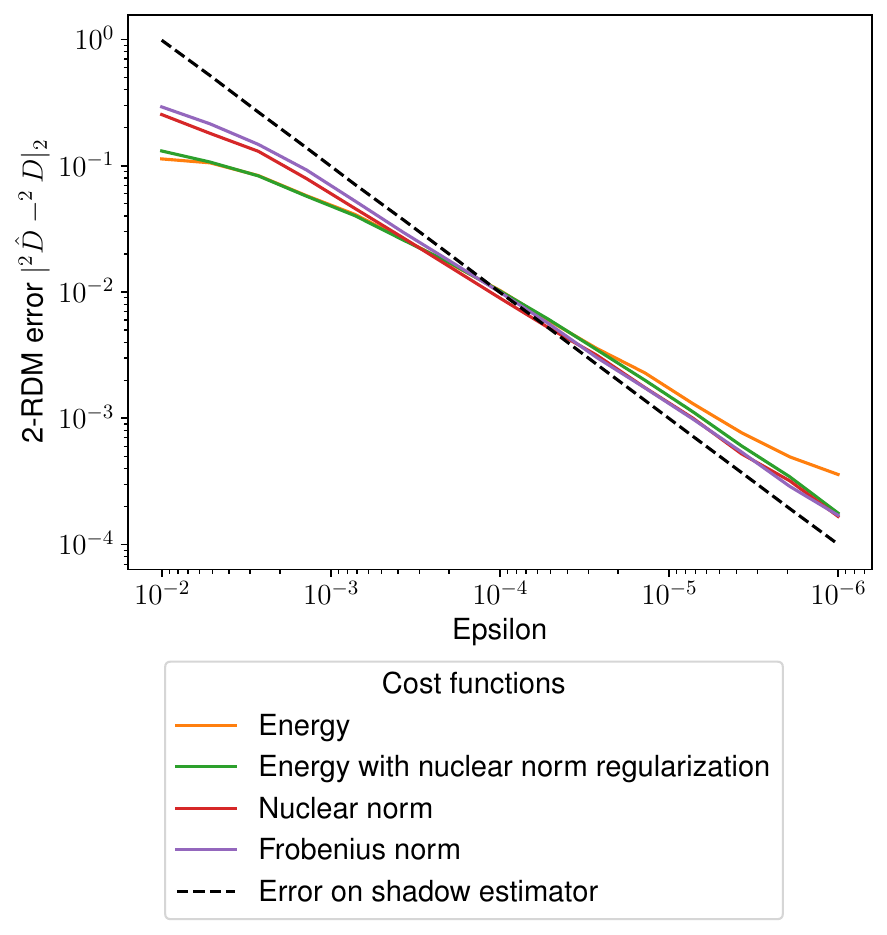}
    \caption{Comparison of different choices of cost functions described in Equations \ref{eq:nrg_cost}, \ref{eq:nrg_nuc_cost}, \ref{eq:nuc_cost}, \ref{eq:fro_cost} against the plain error on the shadow estimator entering the constraints. The constraints otherwise are kept the same, points are averaged over multiple runs with different errors on the shadow constraints.}
    \label{fig:cost_funcs}
\end{figure}

The constraints are the same for all the cost functions. The optimization is implemented by using the package CVXPY ~\cite{diamond16:_cvxpy} and plotted in Figure \ref{fig:cost_funcs}. The test system is the $(10,8)$ active space for \ce{N2} in the cc-pvdz basis set used in the main text as a testing system. The 2-RDM is cast into a matrix form via a superindex summing $p,q$ and $r,s$ as is already done when setting up the constraints in v2rdm.

We do not find a significant overall difference between the different cost functions. In the noisy, fairly unconstrained setting that extrapolates to the plain v2rdm algorithm the cost functions using energy and the regularized energy clearly dominate, as no real information about the Hamiltonian or its ground state enter the optimization at a certain noise level. In the fairly constrained setting the cost functions only using rank seem to introduce less of a bias over the shadow estimator but still ending up less accurate as the shadow estimator in the high precision regime.

So at least from this small numerical experiment it does not seem that extracting approximate low-rank 2-RDMs from a noisy classical shadow estimation seems to improve over the optimization under energy or the pure shadow estimator.

\section{Constraining with ground state energy}
\label{app:energy_constr}

In the fault tolerant regime, quantum phase estimation provides access to the ground state energy under controlled approximation under certain conditions at reasonable cost. Although providing access to the ground state energy, reduced density matrices are still prohibitively expensive when looking to estimate further molecular observables like e.g. forces for geometric optimization. Therefore one can pose the question, if constraining the energy by the ground state energy in a semidefinite program allows for a more accurate estimation.

\begin{figure}[h]
\includegraphics[width=.4\textwidth]{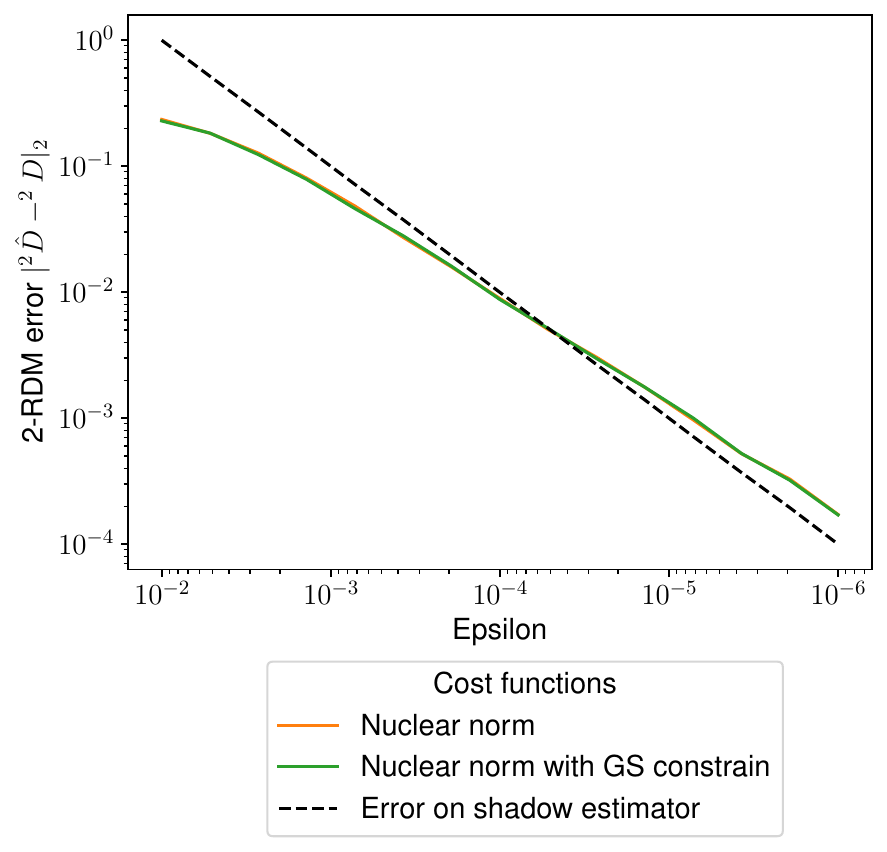}
    \caption{Comparison of optimizing the nuclear norm of cost functions described in Equation \ref{eq:nuc_cost} when omitting/adding the energy constraint from Equation \ref{eq:gs_constr} against the plain error on the shadow estimator. Points are averaged over multiple runs.}
    \label{fig:cost_nrg_constr}
\end{figure}

To include the estimate of the ground state energy in our semidefinite program, we add the following constraint to the optimization 
\begin{equation}
     E_\text{GS} -\epsilon_\text{GS} \leq \operatorname{Tr} (H \,^2\mathbf{D}) \leq E_\text{GS} +\epsilon_\text{GS} ,
     \label{eq:gs_constr}
\end{equation}
where $E_\text{GS}$ is the ground state energy (calculated by FCI in PySCF) and $\epsilon_\text{GS} = 10^{-3}$ Hartree at chemical accuracy, the precision a quantum phase estimation estimate would aim to achieve. The test system is again the $(10,8)$ active space for \ce{N2} in the cc-pvdz basis set used in the main text. Results are plotted in Figure \ref{fig:cost_nrg_constr}.
In our limited data set we do not find an improvement over not including the ground state energy in our optimization. Adding this constraint does not seem to be restricting our allowed solution space in a meaningful way.
\end{document}